\documentclass[%
 preprint,
showpacs,preprintnumbers,
 amsmath,amssymb,
 aps,
]{revtex4-1}

\usepackage{graphicx}

\newcommand{\myfig}[4][ht]{
\begin{figure}[#1]
\centering
\includegraphics[#2]{#3}
\caption{#4\label{#3}}
\end{figure}
}

\newcommand{\myfigwide}[4][ht]{
\begin{figure*}[#1]
\centering
\includegraphics[#2]{#3}
\caption{#4\label{#3}}
\end{figure*}
}

\newcommand{\npar}{%
\\[-3mm]\par}

\usepackage{dcolumn}
\usepackage{bm}

\begin{document}

\title{Nonlocality in microscale heat conduction}


\author{Bjorn Vermeersch}
\email{bvermeer@purdue.edu}
\author{Ali Shakouri}
\email{shakouri@purdue.edu}
\affiliation{\vspace{3mm}Birck Nanotechnology Center, Purdue University, West Lafayette, Indiana 47907, USA}
\date{\today}

\begin{abstract}
Thermal transport at short length and time scales inherently constitutes a nonlocal relation between heat flux and temperature gradient, but this is rarely addressed explicitly. Here, we present a formalism that enables detailed characterisation of the delocalisation effects in nondiffusive heat flow regimes. A convolution kernel $\kappa^{\ast}$, which we term the nonlocal thermal conductivity, fully embodies the spatiotemporal memory of the heat flux with respect to the temperature gradient. Under the relaxation time approximation, the Boltzmann transport equation formally obeys the postulated constitutive law and yields a generic expression for $\kappa^{\ast}$ in terms of the microscopic phonon properties. Subsequent synergy with stochastic frameworks captures the essential transport physics in compact models with easy to understand parameters. A fully analytical solution for $\kappa^{\ast}(x')$ in tempered L\'evy transport with fractal dimension $\alpha$ and diffusive recovery length $x_{\text{R}}$ reveals that nonlocality is physically important over distances $\sqrt{2-\alpha} \,\,x_{\text{R}}$. This is not only relevant to quasiballistic heat conduction in semiconductor alloys but also applies to similar dynamics observed in other disciplines including hydrology and chemistry. We also discuss how the previously introduced effective thermal conductivity $\kappa_{\text{eff}}$ inferred phenomenologically by transient thermal grating and time domain thermoreflectance measurements relates to $\kappa^{\ast}$. Whereas effective conductivities depend on the experimental conditions, the nonlocal thermal conductivity forms an intrinsic material property. Experimental results indicate nonlocality lengths of 400$\,$nm in Si membranes and $\simeq 1\,\mu$m in InGaAs and SiGe, in good agreement with typical median phonon mean free paths.
\end{abstract}
\maketitle
\section{INTRODUCTION}
Microscale heat conduction in semiconductor materials has received rapidly intensifying interest over the past years \cite{cahill,siemens,minnich-spotsize,grating-experiments,johnson,malen,maznev-gratingtheory,MFPspectroscopy,collins,wilson1,minnichBTE,minnich3D,wilson2}. These transport regimes not only have immediate practical relevance for electronic devices \cite{devices} but also provide invaluable clues from which fundamental properties of the microscopic heat carriers can be reconstructed \cite{MFPspectroscopy,alloys-part1,alloys-part2}.
\par
In regular diffusive transport, Fourier's famous law relates the heat flux $\vec{q}$ to the local temperature gradient $\vec{\nabla}T$ through the medium's thermal conductivity $\kappa_0$. In the case of one-dimensional heat flow, we have
\begin{equation}
q(x,t) = -\kappa_0 \frac{\partial T}{\partial x}(x,t) \label{FGR0}
\end{equation}
However, at length and/or time scales comparable to phonon mean free paths (MFPs) and/or relaxation times, the Fourier law breaks down, and the transport no longer behaves as predicted by standard diffusive theory. Prime observations of such `quasiballistic' regimes have been achieved through time domain thermoreflectance (TDTR) \cite{cahill,minnich-spotsize,alloys-part2} and transient thermal grating (TTG) \cite{grating-experiments,johnson} experiments. A growing body of work has been undertaken to develop a deeper understanding of the underlying physics \cite{maznev-gratingtheory,collins,wilson1,minnichBTE,minnich3D,wilson2,alloys-part1,alloys-part2}.
\par
So far, most studies have focused on the anomalous shape or temporal signature of the thermal field. Far fewer works have accounted for the fact that the quasiballistic heat flux in a given location at a given time inherently co-depends on the temperature gradient in other places and/or earlier times \cite{nonlocal1,nonlocal2,nonlocal3}. Put differently, the flux-gradient relation (FGR) or `constitutive law' is no longer localised in nondiffusive transport regimes, but instead possesses `memory' in space and/or time. Building upon the pioneering work by Mahan and Claro for steep gradients \cite{nonlocal1}, Koh and coworkers recently proposed a nonlocal heat conduction theory \cite{kohnonlocal} to investigate quasiballistic transport observed in TDTR experiments on semiconductor alloys \cite{cahill}. Although this work enabled validation of the main trends, the theory explicitly relied on a set of simplifying assumptions in an ideal Debye crystal and consequently only applies to `weakly quasiballistic' regimes in the high temperature limit. Moreover, the nonlocal heat equation had to be solved numerically with a finite element scheme that required several thousands of nodes, and lacked flexibility for direct comparison with actual measurement results.
\par
In this work, we provide a nonlocal formalism that can describe thermal transport regimes across the entire ballistic-diffusive spectrum without imposing any a priori assumptions on the phonon population of the conducting medium. We demonstrate that the Boltzmann transport equation under the relaxation time approximation formally obeys the postulated constitutive law and derive its spatiotemporal FGR memory kernel. The essential physics of quasiballistic transport regimes can then be captured in highly compact form through synergy with stochastic frameworks. TTG and TDTR measurements yield nonlocality distances on the order of typical median phonon MFPs, as intuitively expected.
\section{THEORETICAL FORMALISM}
\subsection{Nonlocal framework}
We assume a homogeneous, isotropic medium and perform the derivation for one-dimensional heat flow. A 1D framework will suffice for our purposes, as the dominant physics relevant to TTG and TDTR experiments can be understood in terms of in-plane and cross-plane transport respectively. We note that explanation of laser spot size effects \cite{minnich-spotsize} does require detailed 3D analysis \cite{minnich3D} and will not be attempted here. In nondiffusive conduction regimes the conventional Fourier law (\ref{FGR0}) no longer holds. Instead, we can postulate a constitutive law of the form
\begin{equation}
q(x,t) = - \int \limits_{0}^{t} \mathrm{d}t' \int \limits_{-\infty}^{\infty} \mathrm{d}x' \kappa^{\ast}(x-x',t-t') \frac{\partial T}{\partial x}(x',t') \label{convolution}
\end{equation}
Here, $\kappa^{\ast}(x',t')$ embodies the spatial and temporal nonlocality of the heat flux with respect to the temperature gradient. We will term this quantity the `nonlocal thermal conductivity kernel' of the medium. We immediately stress that $\kappa^{\ast}$ must not be confused with the `effective thermal conductivity' $\kappa_{\text{eff}}$ commonplace in virtually all quasiballistic heat experiments \cite{cahill,siemens,minnich-spotsize,johnson,grating-experiments,malen}. The latter properties arise through phenomenological interpretations $q(x,t) = -\kappa_{\text{eff}}(x,t) \, \partial_x T(x,t)$ that thus maintain a fully localised FGR, in sharp contrast with (\ref{convolution}). Moreover, $\kappa_{\text{eff}}$ varies with experimental conditions and becomes, by construction, a function of grating period (TTG) or pump modulation (TDTR), whereas $\kappa^{\ast}$ acts as an intrinsic material property of the conducting medium.
\par
Qualitatively, the emergence of FGR delocalisation can be easily understood from microscopic heat conduction fundamentals. By definition, quasiballistic heat conduction in semiconductors is governed by phonons whose mean free path and/or relaxation time are comparable to the spatial and/or temporal temperature gradient. These heat carriers then indeed forge long-range connections between portions of the thermal field that are diffusively speaking far apart in space and/or time.
\par
On the quantitive side, (\ref{convolution}) represents a double convolution and thus greatly simplifies by reverting to spatial and temporal frequency domains. After Fourier ($x \leftrightarrow \xi$) and Laplace ($t \leftrightarrow s$) transformations we have
\begin{equation}
q(\xi,s) = - j \xi \, \kappa^{\ast}(\xi,s) \, T(\xi,s) \label{qP1}
\end{equation}
with $j$ the complex unit. In a medium with volumetric heat capacity $C_0$, conservation of energy requires that $- \partial_x q = C_0 \, \partial_t T$ at all places and times, and hence
\begin{equation}
-j \xi q(\xi,s) =  s P(\xi,s) - P(\xi,t=0) \label{qP2}
\end{equation}
where we introduced the volumetric energy density $P = C_0 \, T$. Both $T$ and $P$ express deviations relative to the ambient background. We now consider a planar source at $x=0$ inside the medium that injects 1$\,$J/m$^2$ at time zero to find the single pulse energy density response $\mathcal{P}(\xi,s)$ from which all dynamic system properties can be derived. Given the initial condition $\mathcal{P}(x,t=0) = \delta(x)$ we have $\mathcal{P}(\xi,t=0) = 1$, so combining (\ref{qP1}) and (\ref{qP2}) produces
\begin{equation}
\kappa^{\ast}(\xi,s) = \frac{C_0}{\xi^2} \left[ \frac{1}{\mathcal{P}(\xi,s)} - s \right] \label{kappamacro}
\end{equation}
Our derivation solely relied on energy conservation and basic isotropy, without making any other assumptions about the medium's crystal configuration or phonon characteristics. The result (\ref{kappamacro}) is therefore universally applicable to all transport regimes across the ballistic-diffusive spectrum in a wide variety of materials.
\par
We can show that the 1D Boltzmann transport equation (BTE) under the relaxation time approximation (RTA) is formally compatible with the delocalised FGR postulated in (\ref{convolution}). Inserting the analytical single pulse response $\mathcal{P}(\xi,s)$ of the RTA-BTE, derived in detail elsewhere \cite{alloys-part1}, into (\ref{kappamacro}) yields
\begin{equation}
\kappa^{\ast}(\xi,s) = \left( \sum \limits_{k} C_k \right) \cdot \left( \sum \limits_{k} \frac{\kappa_k}{(1 + s\tau_k)^2 + \xi^2 \Lambda_{\parallel,k}^2} \right) \biggr/ \left(\sum \limits_{k} \frac{C_k (1+s \tau_k)}{(1 + s \tau_k)^2 + \xi^2 \Lambda_{\parallel,k}^2} \right) \label{kappamicro}
\end{equation}
This result expresses the `macroscopic' nonlocal conductivity kernel $\kappa^{\ast}$ in terms of the underlying microscopic phonon dynamics. Here, $\kappa_k = \Lambda_{\parallel,k} v_{\parallel,k} C_k$ is the thermal conductivity of a phonon mode resolved for wavevector $\vec{k}$, with mean free path (MFP) $\Lambda_k = v_k \tau_k$, group velocity $v_k$, volumetric heat capacity $C_k$ (unit J/m$^3$-K) and relaxation time $\tau_k$. Subscripts $_{\parallel}$ indicate $x$-projection onto the thermal transport axis. The bulk thermal diffusivity $D_0 = \kappa_0/C_0$ is given by $\sum_k \kappa_k / \sum_k C_k$. The summations running over a discrete wavevector grid (and, implicitly, over phonon branches) in (\ref{kappamicro}) enable seamless analysis of \textit{ab initio} phonon data, but can be easily converted to continuous integrals over wavevector space or phonon frequency for analytical modelling. In particular, for an idealised crystal with lattice constant $a$ and spherically symmetric Brillouin zone (BZ) we have
\begin{multline}
\kappa^{\ast}_{\text{SBZ}}(\xi,s) = C_0 \left( \frac{1}{(2\pi)^2} \int \limits_{0}^{\pi} \sin \theta \, \mathrm{d}\theta \int \limits_{0}^{\pi/a} k^2 \mathrm{d}k \, \frac{\Lambda(k) \, v(k) \, C(k)\cos^2 \theta}{[1 + s \tau(k)]^2 + \xi^2 \Lambda^2(k) \cos^2 \theta} \right) \\
\biggr/ \left( \frac{1}{(2\pi)^2} \int \limits_{0}^{\pi} \sin \theta \, \mathrm{d}\theta \int \limits_{0}^{\pi/a} k^2 \mathrm{d}k \, \frac{C(k) \, [1+s \tau(k)]}{[1 + s \tau(k)]^2 + \xi^2 \Lambda^2(k) \cos^2 \theta} \right) \label{kappamicro-spherical}
\end{multline}
where $C_0 = (1/2\pi^2) \int_{0}^{\pi/a} C(k) k^2 \mathrm{d}k$. We note that in integral formulations $C(k)$ denotes the heat capacity of a single mode (unit J/K and equal to the Boltzmann constant in the high temperature limit). The integrations over the polar angle $\theta$ can be carried out analytically and we find
\begin{equation}
\kappa^{\ast}_{\text{SBZ}}(\xi,s) = C_0 \cdot \frac{\int \limits_{0}^{\pi/a} \frac{k^2 \, \Lambda(k) \, v(k) \, C(k)}{\xi^3 \Lambda^3(k)} \left[ \xi \Lambda(k) - \left[1+s \tau(k) \right] \mathrm{arctan} \left( \frac{\xi \Lambda(k)}{1+s\tau(k)} \right) \right] \mathrm{d}k}{\int \limits_{0}^{\pi/a} \frac{k^2 \, C(k)}{\xi \Lambda(k)} \, \mathrm{arctan} \left( \frac{\xi \Lambda(k)}{1+s\tau(k)} \right) \mathrm{d}k} \label{kappamicro-spherical2}
\end{equation}
\par
The generic expressions (\ref{kappamicro}) and (\ref{kappamicro-spherical2}) always produce a well behaved nonlocal conductivity kernel for which a meaningful Fourier-Laplace inversion can be performed. This indicates that thermal transport governed by the RTA-BTE can indeed be formally understood and analysed in terms of a delocalised FGR. Note that when $\xi \Lambda \ll 1$ and $s\tau \ll 1$, we always find $\kappa^{\ast}(\xi,s) \simeq \kappa_0$, and hence $\kappa^{\ast}(x',t') \simeq \kappa_0 \, \delta(x') \delta(t')$. Thus, for transport occurring over length and time scales that are sufficiently larger than phonon MFPs and relaxation times, the FGR described by (\ref{convolution}) becomes localised again and reduces to Fourier's law, as appropriate. 
\subsection{Earlier literature as special cases}
It is instructive to show that the nonlocal model developed by Koh \textit{et al.} \cite{kohnonlocal} formally correponds to a simple special case of the general framework introduced here. The theory in Ref. \cite{kohnonlocal} was constructed under the explicit assumptions that (i) measured transients are slow compared to phonon relaxation times ($s\tau \ll 1$); (ii) the medium is an ideal Debye crystal (each phonon mode propagates at the sound velocity $v_S$) with spherical BZ; and (iii) the medium operates under its high temperature regime (all phonon energies obey $\hbar \omega \ll k_B T_{0}$ and consequently $C(k) \simeq k_B$ where $k_B$ is the Boltzmann constant and $T_{0}$ the absolute ambient temperature). Calculations were performed in a semi-infinite geometry, which required special care to `reflect' backtravelling phonons reaching the top surface back into the medium. This complication can be avoided by symmetric extension to a fully infinite medium with heat source at $x=0$ as we have assumed earlier. The constitutive law derived in Ref. \cite{kohnonlocal} then reads
\begin{equation}
q_{\text{Koh}}(x,t) = - \frac{k_B v_S}{8 \pi^2} \int \limits_{-\infty}^{\infty} \mathrm{d}x' \frac{\partial T}{\partial x}(x',t) \int \limits_{0}^{\pi} \sin \theta \, |\cos \theta| \, \mathrm{d}\theta \int \limits_{0}^{\pi/a} k^2 \mathrm{d}k \exp \left( - \left| \frac{x-x'}{\Lambda(k) \cos \theta}\right|\right) \label{FGRkoh}
\end{equation}
This obeys the form postulated in our Eq. (\ref{convolution}), with generalised conductivity kernel
\begin{eqnarray}
\kappa^{\ast}_{\text{Koh}}(x',t') & = & \frac{k_B v_S \delta(t')}{8 \pi^2} \int \limits_{0}^{\pi} \sin \theta \, |\cos \theta| \, \mathrm{d}\theta \int \limits_{0}^{\pi/a} k^2 \mathrm{d}k \exp \left( - \left| \frac{x'}{\Lambda(k) \cos \theta}\right|\right) \\
\Rightarrow \kappa^{\ast}_{\text{Koh}}(\xi) & = & \frac{k_B v_S}{(2 \pi)^2} \int \limits_{0}^{\pi} \sin \theta \, \mathrm{d}\theta \int \limits_{0}^{\pi/a} k^2 \mathrm{d}k \, \frac{\Lambda(k) \cos^2 \theta}{1 + \xi^2 \Lambda^2(k) \cos^2 \theta} \label{kappaKoh}
\end{eqnarray}
Note that the final result is independent of $s$ since (\ref{FGRkoh}) is only delocalised in space.
\npar
We now apply the same set of assumptions $s \tau(k) \ll 1, v(k) \equiv v_S, C(k) \simeq k_B$ adopted by Ref. \cite{kohnonlocal} to our convolution kernel (\ref{kappamicro-spherical}) for a spherical BZ crystal. With $C(k)$ being constant, the denominator of $\kappa^{\ast}_{\text{SBZ}}$ is dominated by phonon modes with large density of states, i.e. those towards the outer regions of the BZ. These modes have MFPs that are significantly smaller than typical length scales of experimentally induced thermal gradients, and we can safely assume $\xi \Lambda(k) \ll 1$ for this part of the calculation. Consequently the denominator simply tends to the bulk volumetric heat capacity $C_0$, and (\ref{kappamicro-spherical}) reduces to
\begin{equation}
\kappa^{\ast}_{\text{SBZ}}(\xi) \simeq \frac{k_B v_S}{(2 \pi)^2} \int \limits_{0}^{\pi} \sin \theta \, \mathrm{d}\theta \int \limits_{0}^{\pi/a} k^2 \mathrm{d}k \, \frac{\Lambda(k) \cos^2 \theta}{1 + \xi^2 \Lambda^2(k) \cos^2 \theta} \label{kappaSBZpoisson}
\end{equation}
which is indeed identical to (\ref{kappaKoh}). Contrary to previous work, however, the framework introduced here is equally applicable to fully ballistic ($s\tau \geq 1$) and low temperature regimes, and can readily handle arbitrary BZ geometries and phonon dispersions.
\par
Several other studies have sought to extend the Fourier law to quasiballistic heat flow regimes through harmonic series analysis of BTE solutions for grey or dual phonon channel media \cite{constitutive1,constitutive2}. The resulting heat flux remains fully localised, but involves a correction term in $\partial^3 T / \partial x^3$. It can be shown that these previously derived constitutive laws unknowingly but rigorously correspond to the second order Taylor series expansion of the generalised conductivity kernel $\kappa^{\ast}$ with respect to spatial frequency $\xi$. Full details will be discussed in a separate publication.
\subsection{Synergy with stochastic frameworks}
Stochastic transport models, which describe the governing dynamics in terms of random motion of energy carriers, have been well established in the literature for a wide variety of applications \cite{klafter1,klafter2,CTRWmaster2}. We briefly review the key fundamentals relevant to our discussion of quasiballistic heat conduction. Within the stocastic context, the single pulse response $\mathcal{P}(x,t)$ expresses the chance to find a randomly wandering energy packet in location $x$ at time $t$ after it was released by the source in $x=0$ at $t=0$. In the case of 1D flight processes, the packet moves through consecutive jumps $x(t+\vartheta) = x(t) + \zeta$ where the jump length $\zeta$ and wait time $\vartheta$ are randomly chosen from stochastically independent distributions $\phi(\zeta)$ and $\varphi(\vartheta)$. The Montroll-Weiss equation \cite{CTRWmaster1,CTRWmaster2} provides a closed form solution in Fourier-Laplace domain for the single pulse response induced by the flight process:
\begin{equation}
\mathcal{P}(\xi,s) = \frac{1 - \varphi(s)}{s [1 - \phi(\xi) \varphi(s)]} = \frac{\Psi(s)}{s \left[ \Psi(s) + \psi(\xi) \right]} \label{Pflights}
\end{equation}
where we introduced $\psi(\xi) = 1-\phi(\xi)$ and $\Psi(s) = 1/\varphi(s) - 1$ for notation convenience. The associated generalised conductivity kernel given by (\ref{kappamacro}) then reads
\begin{equation}
\kappa^{\ast}(\xi,s) = C_0 \cdot \frac{\psi(\xi)/\xi^2}{\Psi(s)/s} \label{kappaflights}
\end{equation}
In `Poissonian' flight processes $\Psi(s) = s\vartheta_0$ (the wait time distribution is exponential with average $\vartheta_0$), the single pulse response (\ref{Pflights}) and generalised conductivity (\ref{kappaflights}) simplify to
\begin{equation}
\text{Poissonian:} \quad \mathcal{P}(\xi,s) = \frac{1}{s + \psi(\xi)/\vartheta_0} \quad , \quad \kappa^{\ast}(\xi) = C_0 \, \frac{\psi(\xi)}{\vartheta_0 \, \xi^2} \label{kappapoisson}
\end{equation}
The $s$ dependence has vanished from $\kappa^{\ast}$: the FGR in Poissonian transport has no temporal memory. The FGR memory kernel (\ref{kappamicro}) of the RTA-BTE becomes equally insensitive to $s$ whenever $s\tau \ll 1$. Therefore, `weakly quasiballistic' transport as termed by Hua and Minnich \cite{minnichBTE} is inherently Poissonian and only delocalised in space. Relaxation times of the dominant heat carriers in semiconductors typically span the sub-ps to 2--3$\,$ns range \cite{alloys-part1}, with the longest living modes having the smallest density of states. We can therefore reasonably assume that the Poissonian regime extends up to $s \tau_{\text{max}} \leq 1$, or roughly $f \leq 50\,$MHz, and thus encompasses the typical operational range of TTG and TDTR experiments \cite{minnichBTE,kohnonlocal}.
\par
Several subtypes of Poissonian flight processes deserve specific mentioning. The first consists of regular Fourier diffusion, whose single pulse response $\mathcal{P}(\xi,s) = 1/(s+D_0 \xi^2)$ corresponds to $\psi(\xi) = L_0^2 \, \xi^2$. The characterstic jump length $L_0$ and wait time $\tau_0$ jointly govern the bulk diffusivity through $L_0^2/\vartheta_0 = D_0$. From (\ref{kappapoisson}) we have $\kappa^{\ast}(\xi,s) \equiv \kappa_0$ so the FGR correctly reduces to Fourier's law. Second, $\psi(\xi) = L_0 |\xi|$ describes a ballistic regime with Cauchy-Lorentz distributions $\mathcal{P}(\xi,s) = 1/(s+v_0 |\xi|)$ and $\kappa^{\ast}(\xi) = C_0 v_0/ |\xi|$ where $v_0 = L_0/\vartheta_0$ is the average propagation velocity. Finally, intermediate cases $\psi(\xi) = L_0^{\alpha} \, |\xi|^{\alpha}$ ($1 < \alpha < 2$) correspond to L\'evy superdiffusion with fractal dimension $\alpha$ and fractional diffusivity $D_{\alpha} = L_0^{\alpha}/\vartheta_0$ (unit m$^{\alpha}$/s). The single pulse response is governed by alpha-stable distributions which have signature `heavy tails' $\mathcal{P}(x \gg [D_{\alpha} t]^{1/\alpha},t) \sim |x|^{-(\alpha+1)}$. Now $\kappa^{\ast}(\xi) = C_0 D_{\alpha}/|\xi|^{2-\alpha}$ so the spatial heat flux memory decays algebraically with distance: $\kappa^{\ast}(x') \sim |x'|^{-(\alpha-1)}$. The characteristic blueprints for the three discussed cases are summarised in Table \ref{tab:kappapoisson}.
\begin{table}[!htb]
\caption{Nonlocal thermal conductivity kernel $\kappa^{\ast}(\xi)$ for Poissonian thermal transport in isotropic medium with bulk conductivity $\kappa_0$ and volumetric heat capacity $C_0$.}\label{tab:kappapoisson}
\vspace{5mm}
\begin{tabular}{ccc}
\hline
\textbf{Transport regime} & \textbf{Governing Parameter(s)} & $\kappa^{\ast}(\xi)$ \textbf{[W/m-K]}\\
\hline
Fourier diffusion & bulk diffusivity $D_0$ & $C_0 D_0 = \kappa_0$ \\
L\'evy superdiffusion & fractal dimension $1<\alpha<2$ & \\
& fractional diffusivity $D_{\alpha}$ & $C_0 D_{\alpha}/|\xi|^{2-\alpha}$\\
Ballistic (Cauchy-Lorentz) & avg. propagation velocity $v_0$ & $C_0 v_0 / |\xi|$ \\
\hline
\end{tabular}
\end{table}
\section{RESULTS}
\subsection{Spatiotemporal heat flux memory in common semiconductors}
Using ab initio phonon populations courtesy of J. Carrete and N. Mingo \cite{alloys-part1} as input to Eq. (\ref{kappamicro}), we evaluate the theoretical nonlocal conductivity kernel in Si, InGaAs and SiGe.
\myfigwide[!htb]{width=\textwidth}{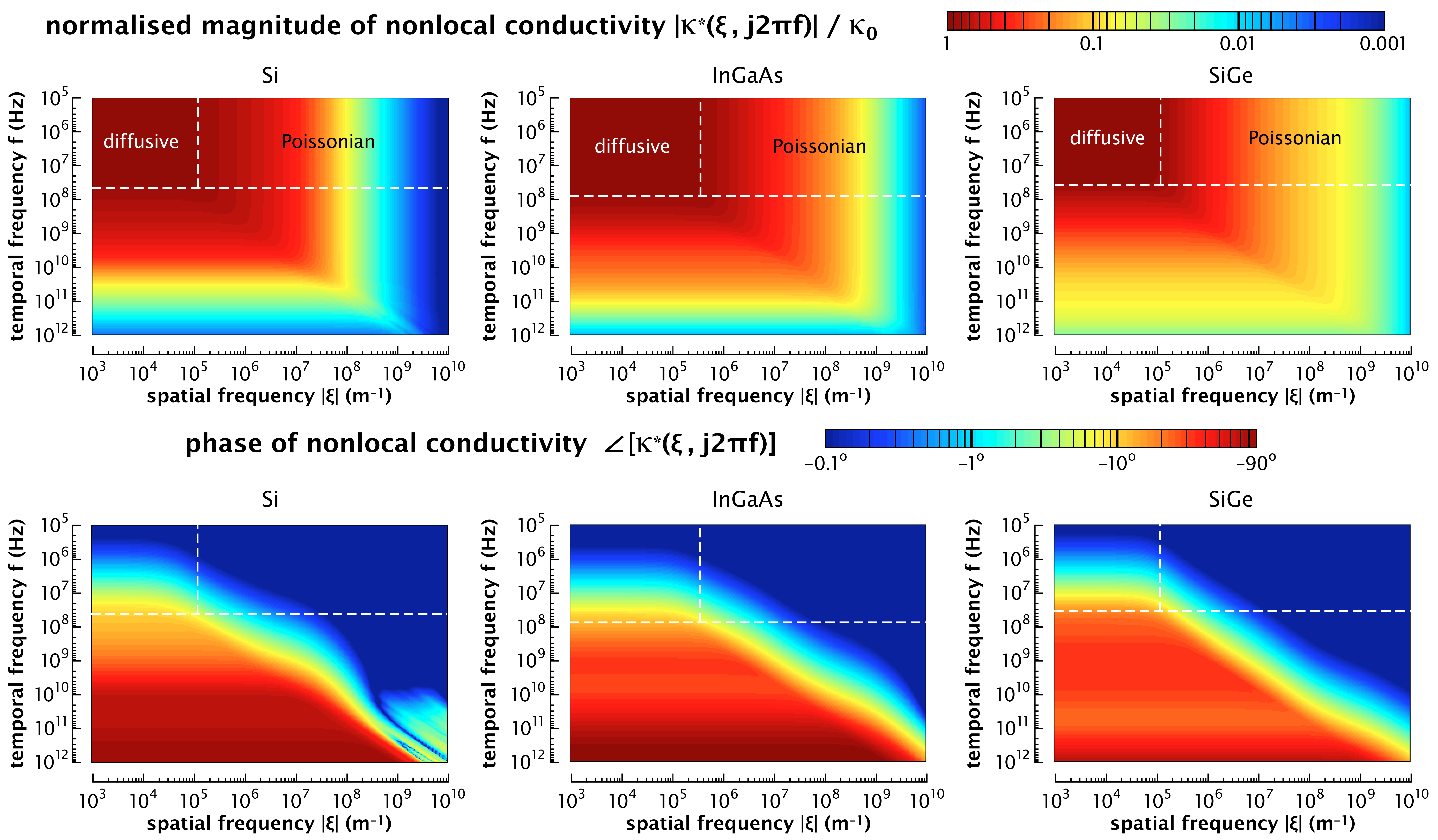}{Nonlocal thermal conductivity kernel $\kappa^{\ast}(\xi,s=j2\pi f)$ evaluated from Eq. (\ref{kappamicro}) with ab initio phonon populations for Si, InGaAs and SiGe. The purely diffusive and Poissonian quasiballistic regimes can be easily identified in the magnitude plots.}
\myfig[!htbw]{width=0.4\textwidth}{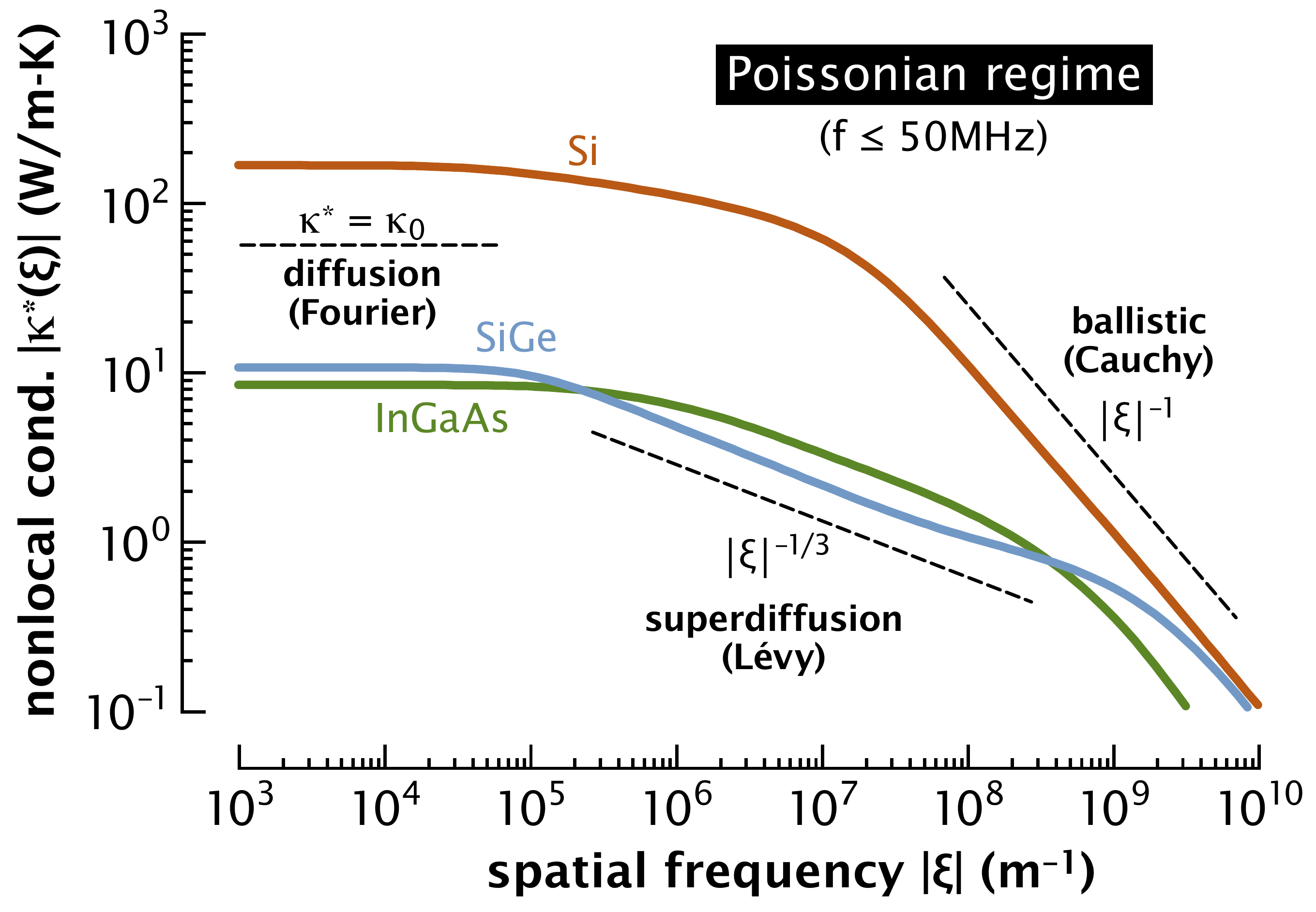}{Nonlocal thermal conductivity kernel $\kappa^{\ast}(\xi)$ for Poissonian transport, i.e. transients that are slow compared to phonon relaxation times, calculated from ab initio phonon populations. Distinct regimes are easily identified using Table \ref{tab:kappapoisson}. The emergence of fractal L\'evy dynamics with $\alpha \simeq 5/3$ in alloy materials is clearly visible.}
\par
Full transient solutions $\kappa^{\ast}(\xi,s)$ (Fig. 1) become virtually independent of temporal frequency below the 40--80$\,$MHz range, in good agreement with the 50$\,$MHz Poissonian threshold anticipated earlier. From the Poissonian limit $\kappa^{\ast}(\xi,s \rightarrow 0)$ (Fig. 2) we can easily identify distinct regimes based on the characteristic signatures from Table I. In Si we observe a fairly smooth, continuous transition from ballistic transport at short length scales to regular diffusion at long length scales. InGaAs and SiGe, by contrast, exhibit an additional intermediate regime where $\kappa^{\ast} \sim \xi^{-1/3}$, indicating L\'evy transport with fractal dimension $\alpha \simeq 5/3$. The emergence of L\'evy dynamics in semiconductor alloys has only been pointed out very recently \cite{alloys-part1,alloys-part2} and physically originates in their large phonon scattering exponent. The behaviour can be better understood by recalling the nonlocal conductivity kernel (\ref{kappaSBZpoisson}) for Poissonian transport in a Debye crystal with spherical BZ. To simplify the analysis, we replace $\cos^2 \theta$ (where $\theta$ sweeps from $0$ to $\pi$) in the $k$ integral by its average value $1/2$ and operate under a full continuum limit (lattice constant $a \rightarrow 0$). This yields
\begin{equation}
\kappa^{\ast}(\xi) \sim \int \limits_{0}^{\infty} \frac{k^2 \, \Lambda(k) \, \mathrm{d}k}{2 + \xi^2 \, \Lambda^2(k)} \label{kappaLevyorigin}
\end{equation}
If we now assume an idealised alloy with phonon scattering mechanism of the form $\Lambda(k) \sim k^{-n}$ ($n>3$, not necessarily integer), (\ref{kappaLevyorigin}) can be evaluated fully analytically:
\begin{equation}
\Lambda(k) \sim k^{-n}: \quad \kappa^{\ast}(\xi) \sim |\xi|^{-(1-3/n)} \quad \Rightarrow \quad \alpha = 1 + \frac{3}{n} \quad (n>3) \label{levyorigin}
\end{equation}
This indeed shows the natural emergence of L\'evy dynamics, and illustrates the intimate connection between the macroscopic fractal dimension $\alpha$ and the microscopic phonon scattering exponent $n$. This derivation also offers semi-analytical validation of the results we previously deduced asymptotically from numerical trends \cite{alloys-part1}.
\subsection{Flux delocalisation in tempered L\'evy transport}
From (\ref{levyorigin}) it is clear that the considered idealised alloy would maintain pure L\'evy transport indefinitely (at all length scales). This is because a purely algebraic scattering relation produces arbitrarily large MFPs near the BZ center $\Lambda(k \rightarrow 0) \rightarrow \infty$ while the continuum limit allows arbitrarily small MFPs $\Lambda(k\rightarrow\infty)\rightarrow 0$. In reality, the MFP spectrum is physically bounded on either side, causing the L\'evy transport to transition to diffusive and ballistic dynamics at respectively long and short length scales as observed in Fig. 2.
\par
A closer investigation of the heat flux delocalisation in realistic semiconductor alloys will need to properly account for the gradual L\'evy-Fourier regime transition. In previous work, we introduced a Poissonian `truncated L\'evy' (TL) formalism for quasiballistic thermal characterisation \cite{alloys-part2}. This model provides accurate fits to raw TDTR measurement data from which we inferred experimental fractal dimensions $\alpha \simeq 1.68$ in InGaAs and SiGe samples, in near perfect agreement with the ab initio predictions. Internally, the TL model is based upon an algebraic jump length distribution whose tail is exponentially suppressed with rate parameter $u_{\text{BD}}$. Physically, $u_{\text{BD}}$ acts as a characteristic length scale over which the macroscopic thermal field achieves the gradual recovery from L\'evy transport to regular diffusion. Using Eq. (7) in Ref. \cite{alloys-part2}, the FGR spatial memory kernel associated with the TL model is readily found to be
\begin{equation}
\kappa^{\ast}(\xi) = \frac{2 \, \kappa_0}{\alpha \, (\alpha-1) \, \xi^2 \, u_{\text{BD}}^2} \left\{ 1 - \left( \xi^2 u_{\text{BD}}^2 + 1 \right)^{\alpha/2} \, \cos \left[ \alpha \, \mathrm{arctan} \left( \frac{1}{|\xi| u_{\text{BD}}} \right) - \frac{\alpha \pi}{2} \right] \right\} \label{kappaxi-TL}
\end{equation}
Once values for $\alpha$ and $u_{\text{BD}}$ are specified, (\ref{kappaxi-TL}) can be transformed to the real space domain using numerical Fourier inversion.
\par
Here, we will follow a slightly different approach that preserves all key physical aspects but greatly simplifies the mathematical treatment. Recalling the functional signatures for Poissonian flights (Table I), we can postulate a nonlocal diffusivity kernel of the form:
\begin{equation}
D^{\ast}(\xi) = \frac{\kappa^{\ast}(\xi)}{C_0} = \frac{D_0}{\left( 1 + x_{\text{LF}}^2 \xi^2 \right)^{1-\alpha/2}} \quad (1 < \alpha < 2) \label{kappaLFxi}
\end{equation}
Over short distances $x_{\text{LF}} \xi \gg 1$ this corresponds to L\'evy transport with fractal dimension $\alpha$ and fractional diffusivity $D_{\alpha} = D_0/x_{\text{LF}}^{2-\alpha}$ while the long range limit $x_{\text{LF}} \xi \ll 1$ exhibits regular diffusion with bulk diffusivity $D_0$. The parameter $x_{\text{LF}}$ thus acts as a characteristic length scale for L\'evy-Fourier recovery, just like its $u_{\text{BD}}$ counterpart. While the asymptotic limits and qualitative behaviour of the two tempered L\'evy models are identical, slight quantitative differences are obviously inevitable. In particular, we point out that for $\alpha \simeq 5/3$ a parameter ratio $u_{\text{BD}}/x_{\text{LF}} \simeq 3.8$ emerges in order to preserve the same $D_{\alpha}$. The experimental characterisations on InGaAs and SiGe performed in Ref. \cite{alloys-part2} then produce $x_{\text{LF}}$ values on the order of one micron, in agreement with typical median phonon MFPs.
\par
It is noteworthy that the stochastically inspired expression (\ref{kappaLFxi}) describes a generic transition from L\'evy transport to regular diffusion without making particular assumptions about the nature of the microscopic carriers. The analysis that follows, therefore, is not limited to heat conduction in semiconductor alloys but can also provide valuable insight to other disciplines where similar dynamics have been observed. Notable examples include sediment transport in rivers \cite{hydrology1,hydrology2} and solute chemical reactions in flows \cite{chemistry1,chemistry2}.
\par
A key advantage of the alternative tempered L\'evy model (\ref{kappaLFxi}) is that the FGR convolution kernel can transformed fully analytically to the real space domain:
\begin{equation}
D^{\ast}(x') = \frac{2^{\nu} D_0}{\sqrt{\pi} \, \Gamma (1/2-\nu) \, x_{\text{LF}}} \cdot \frac{K_{\nu} (|x'/x_{\text{LF}}|)}{|x'/x_{\text{LF}}|^{\nu}} \quad , \quad \nu = \frac{\alpha-1}{2} \label{kappaLF}
\end{equation}
where $K$ is the modified Bessel function of the second kind. The inner core of the kernel exhibits pure L\'evy behaviour $D^{\ast}(|x'| \ll x_{\text{LF}}) \sim |x'/x_{\text{LF}}|^{-(\alpha-1)}$, while the tails decay much more rapidly as $|x'/x_{\text{LF}}|^{-\alpha/2} \exp (- |x'/x_{\text{LF}}|)$. Given that $D^{\ast}(\xi = 0) = D_0$, the total kernel content $\int_{-\infty}^{\infty}D^{\ast}(x') \mathrm{d}x'$ always equals the bulk diffusivity. The tempered L\'evy parameters regulate how this fixed memory budget gets spatially distributed: $x_{\text{LF}}$ sets the overall characterstic length scale, while $\alpha$ governs the shape details. For example, smaller (more superdiffusive) $\alpha$ values produce kernels that are less sharply concentrated near the origin in favour of stronger tails (Fig. 3), consistent with the notion that delocalisation should become more prominent as transport dynamics deviate further from regular diffusion.
\myfig[!htb]{width=0.4\textwidth}{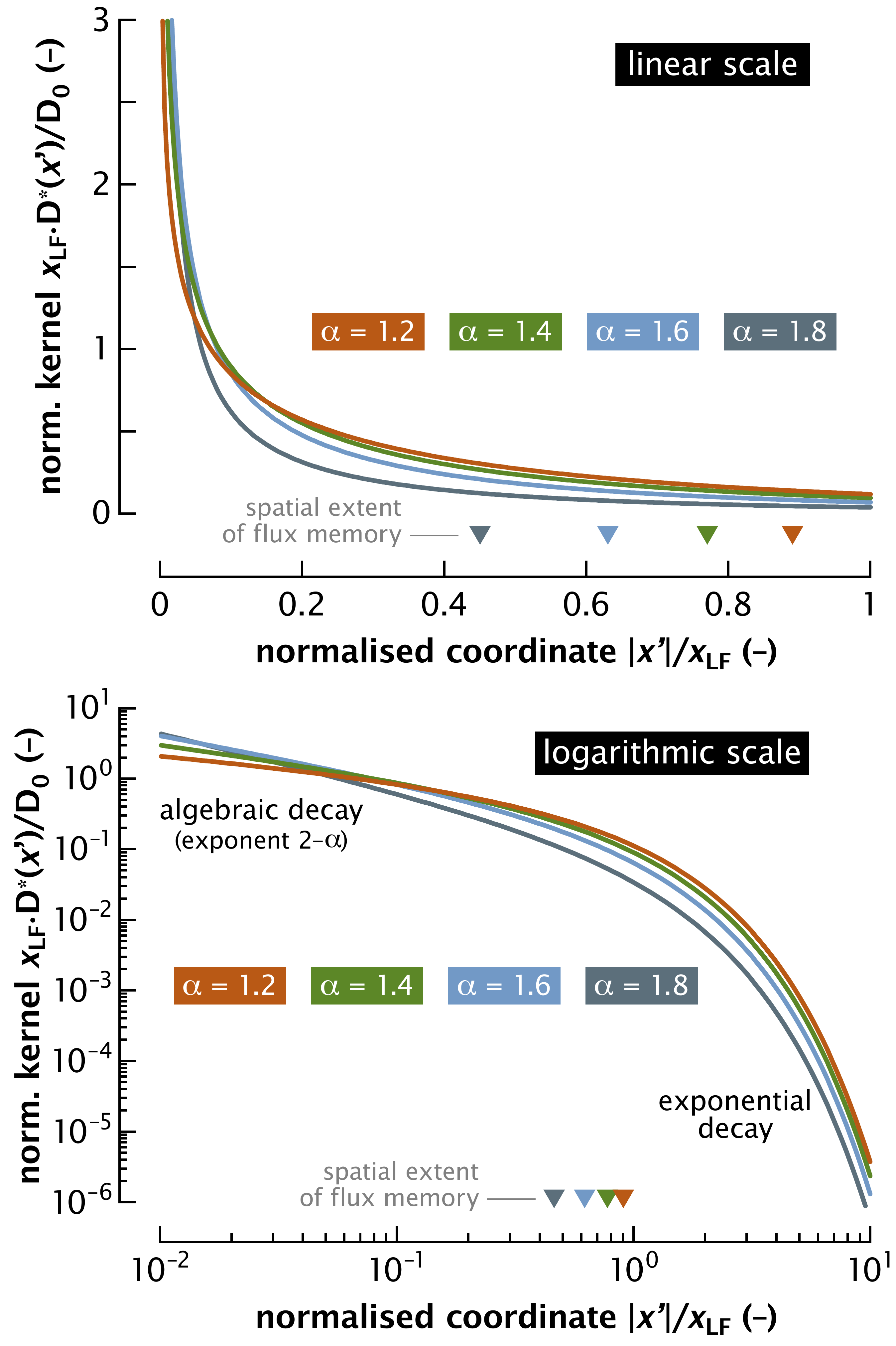}{Dimensionless nonlocal diffusivity kernel in real space domain for tempered L\'evy transport as evaluated from analytical solution (\ref{kappaLF}). $\alpha$ denotes the fractal dimension of the L\'evy regime while $x_{\text{LF}}$ sets the length scale over which the transport recovers to regular diffusion. The characteristic `width' of the kernel, which acts as a measure for the distance over which flux-gradient delocalisation effects are physically important, equals $\sqrt{2-\alpha}\,\,x_{\text{LF}}$ and is indicated by the arrow marks.}
\par
We can easily quantify the associated spatial extent over which nonlocal effects are physically important by considering $D^{\ast}(x')/D_0$. This quantity acts like a properly normalised probability density function for which we analytically obtain a standard deviation $\sqrt{\left< X'^2 \right>} = \sqrt{2-\alpha} \,\, x_{\text{LF}}$. Notice that $\left< X'^2 \right> \rightarrow 0$ in the diffusive limit $\alpha \rightarrow 2$, signaling a Dirac collapse of the kernel and recovery to the localised Fourier/Fick law as appropriate.
\section{CONNECTIONS WITH EXPERIMENTAL CHARACTERISATION}
Most quasiballistic transport observations are interpreted through `modified Fourier' theory. This means that the measurement is analysed with a model that is still purely diffusive but with phenomenologically adjusted bulk thermal conductivity to obtain the best possible fit with the recorded data. Applying this procedure to TTG and TDTR experiments produces the `effective' thermal conductivity of the semiconductor as a function of grating period and pump modulation frequency respectively. In this section, we determine how these effective conductivities relate to the medium's intrinsic nonlocal conductivity kernel.
\subsection{Transient thermal grating}
TTG employs the interference pattern of two laser beams to impose a heat source that is periodic in space \cite{johnson}. This produces a sinusoidal temperature profile with grating period $\lambda$ at the semiconductor surface that then equilibrates predominantly through 1D in-plane transport. TTG thus essentially probes the medium's Fourier domain response at a single spatial frequency $\xi_{\lambda} = 2\pi / \lambda$. In diffusive regime, the peak-to-valley temperature contrast of the grating pattern decays exponentially with time constant $\Theta_{0} = 1/D_0 \, \xi_{\lambda}^2 = \lambda^2/4\pi^2 D_0$ \cite{johnson}.
\par
Grating periods that overlap with phonon MFPs induce nondiffusive transport. Most experiments probe a regime where the BTE solution still decays exponentially, but slower than predicted by diffusive theory \cite{minnichBTE}. Identifying the time constant $\Theta$ of an exponential fit to the measured quasiballistic transient with the diffusive expression $\Theta_0$ provides the `effective' thermal conductivity \cite{johnson}
\begin{equation}
\kappa_{\text{eff}}(\lambda) = \frac{C_0}{\Theta \, \xi_{\lambda}^2} \label{TTGkappaeff}
\end{equation}
We point out that the exponential decay of the measured responses is a direct manifestation of Poissonian transport dynamics. Indeed, at a given spatial frequency $\xi_{\lambda}$ the generic pulse response of a Poissonian flight process (\ref{kappapoisson}) decays exponentially in time with time constant $\Theta =\vartheta_0/\psi(\xi_{\lambda})$. Moreover, from (\ref{kappapoisson}) we further obtain
\begin{equation}
\kappa^{\ast}(\xi_{\lambda}) = \frac{C_0 \, \psi(\xi_{\lambda})}{\vartheta_0 \, \xi_{\lambda}^2} = \frac{C_0}{\Theta \, \xi_{\lambda}^2} \label{TTGkappaxi}
\end{equation}
which is identical to (\ref{TTGkappaeff}). The effective conductivity inferred at grating period $\lambda$ thus formally corresponds to a direct sampling of the FGR memory kernel $\kappa^{\ast}(\xi)$ at spatial frequency $\xi = 2\pi/\lambda$. Put differently, we conclude that in TTG experiments the effective and nonlocal conductivity become fully interchangeable, at least from a 1D perspective. This result can also be understood in terms of Hua and Minnich's finding \cite{minnichBTE} that the 1D BTE solution formally justifies the use of modified Fourier theory for TTG analysis in the weakly quasiballistic (Poissonian) regime.
\par
Although 1D models are sufficient to understand the essential physics of quasiballistic TTG effects \cite{maznev-gratingtheory,collins,minnichBTE}, direct quantitative comparison with raw measurement data is not straightforward. In bulk samples, cross-plane thermal leakage from the surface cannot be neglected, and a 2D diffusion model is typically used to analyse the recorded decay of the thermal field \cite{grating-experiments}. Extension of the nonlocal framework to multidimensional heat flow geometries falls outside the scope of this paper, but may be the subject of future work. Here, we will focus on measurements on Si membranes performed by Johnson and coworkers \cite{grating-experiments}. The submicron thickness and suspended nature of these samples enforces a near ideal 1D thermal field, making them highly suitable for nonlocal heat conduction analysis. Inspired by the Si ab initio results from Fig. 2, we will model nondiffusive transport in the membrane as a gradual transition from a ballistic Cauchy regime to regular Fourier diffusion over characteristic length scale $x_{\text{CF}}$. This can be easily achieved by setting $\alpha=1$ in the tempered L\'evy model used earlier for the alloys:
\begin{equation}
\kappa^{\ast}(\xi) = \frac{\kappa_0}{\sqrt{1 + x_{\text{CF}}^2 \xi^2}} \label{kappaCFxi} \quad
\Leftrightarrow \quad \kappa^{\ast}(x') = \frac{\kappa_0}{\pi x_{\text{CF}}} \, K_0 \left( \frac{|x'|}{x_{\text{CF}}} \right)
\end{equation}
so that we find
\begin{equation}
\kappa_{\text{eff}}(\lambda) = \kappa^{\ast}(\xi = 2\pi/\lambda) = \frac{\kappa_0}{\sqrt{1 + (2\pi x_{\text{CF}}/\lambda)^2}} \label{kappaeffTG}
\end{equation}
Since this expression results from a Poissonian framework, its accuracy should be expected to diminish for very small grating periods as then the characteristic time scale $\Theta \sim \lambda^2$ becomes comparable with phonon relaxation times. This brings temporal flux memory into play, and the non-Poissonian transport corresponds to the `strongly quasiballistic' regime in which the BTE solution can no longer be fitted by a simple exponential decay \cite{minnichBTE}. At room temperature we estimate crossover in typical semiconductors around $\lambda \simeq 1\,\mu$m. The experiments with $\lambda \geq 2\,\mu$m reported in Ref. \cite{grating-experiments} should thus operate within the Poissonian regime, and as a result our simple model indeed performs quite well (Fig. 4).
\myfig[!htb]{width=0.4\textwidth}{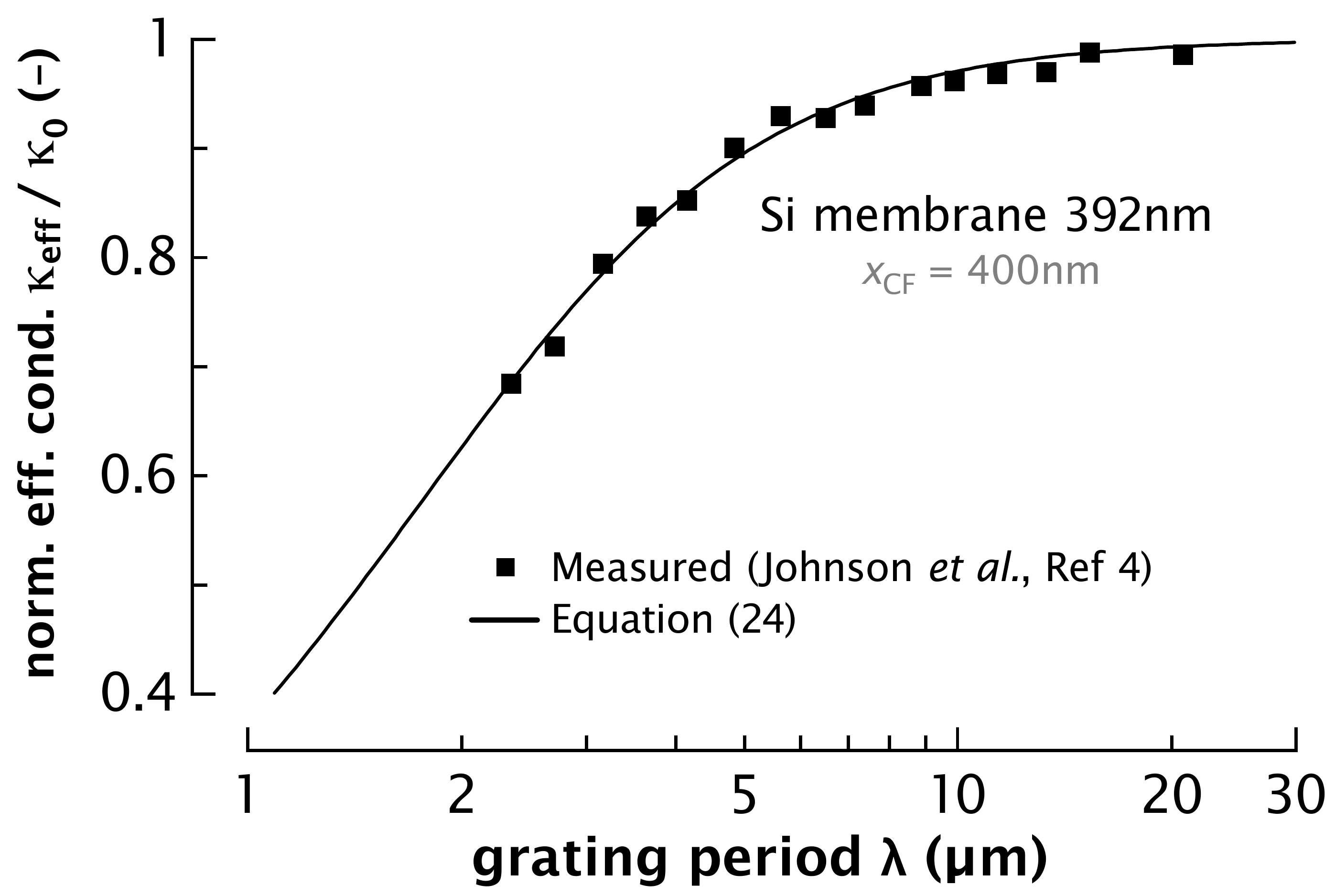}{Thermal transient grating measurement results agree well with our analytical expression (\ref{kappaeffTG}). The $x_{\text{CF}}$ fitting parameter provides the diffusive recovery length scale over which heat flux delocalisation is physically important.}
\par
The best fit is obtained when setting $x_{CF} = 400\,$nm, once again in good agreement with typical median MFPs \cite{MFP2}. We note that Ref. \cite{grating-experiments} used the conductivity of a bulk Si sample as baseline $\kappa_0$ for normalisation of the membrane results. The reported values suggest a Fourier conductivity $\kappa_{\text{eff}}(\lambda \rightarrow \infty) \simeq 0.62 \, \kappa_0$, testament to significant boundary scattering in the very thin film. In Fig. 4 we have applied a uniform scaling factor of $0.62^{-1}\approx 1.61$ to the experimental data for consistent comparison with the analytical model but this obviously does not affect the $x_{\text{CF}}$ identification.
\subsection{Time domain thermoreflectance}
TDTR techniques use laser pulse trains that are modulated at temporal frequency $f_{\text{mod}}$ as heat source \cite{cahill-model,cahill}. The thermal dynamics of the semiconductor can only be studied indirectly since the method requires a thin metal film to be deposited onto the sample surface as measurement transducer. Heat spreading in the transducer, thermal contact resistance of the metal/semiconductor interface and pulse accumulation effects complicate the processing of raw measurement data \cite{cahill-model}. A full 3D truncated L\'evy model described elsewhere \cite{alloys-part2} properly accounts for these issues and enables detailed experimental characterisation of the quasiballistic heat flow in semiconductor alloys.
\par
It is therefore not our aim here to use the nonlocal framework for direct TDTR analysis. Rather, we simply wish to provide some qualitative insight on how the inferred effective thermal conductivity relates to the nonlocal conductivity kernel. While the presence of the transducer and in-plane shape of the laser beams entails great care in the fitting procedures, the information that is ultimately extracted is dominated by 1D cross-plane transport and primarily sensitive to the frequency domain single pulse response $\mathcal{P}(x=0,s=j 2 \pi f_{\text{mod}})$ of the semiconductor surface \cite{cahill-model,kohnonlocal}. For our purposes, we can safely ignore the presence of the transducer and again assume an infinite medium with heat source at $x=0$. In diffusive regime, a modulated heat source induces an exponential temperature gradient inside the medium:
\begin{equation}
Gr(x) = \frac{\partial T}{\partial x}(x) = G_0 \, \exp ( -|x|/\ell) \quad \Leftrightarrow \quad j\xi \, T(\xi) = \frac{2 G_0 \ell}{1 + \xi^2 \ell^2} \label{expongradient}
\end{equation}
The decay length $\ell = \sqrt{D_0/\pi f_{\text{mod}}}$ ranges from $\simeq 1\,\mu$m at 1$\,$MHz to $\simeq 250\,$nm at 20$\,$MHz in typical semiconductor alloys. The considerable overlap with phonon MFPs induces quasiballistic transport that is conventionally interpreted with modified Fourier theory, i.e. a diffusive model with phenomenologically adjusted bulk conductivity. Such analyses therefore still assume an exponential gradient, but with an `effective' decay length $\ell_{\text{eff}}$. Within good approximation, the effective conductivity inferred by the measurement corresponds to
\begin{equation}
\kappa_{\text{eff}} \simeq - \frac{q(x=0)}{Gr(x=0)} = -\frac{1}{\pi G_0} \int \limits_{0}^{\infty} q(\xi) \mathrm{d}\xi
\end{equation}
The measurement operates in the Poissonian regime as determined earlier so the heat flux within the nonlocal framework is given by $q(\xi) = -j\xi \, \kappa^{\ast}(\xi) \, T(\xi)$, and we find
\begin{equation}
\kappa_{\text{eff}} = \frac{2 \ell_{\text{eff}}}{\pi} \int \limits_{0}^{\infty} \frac{\kappa^{\ast}(\xi)\mathrm{d}\xi}{1 + \xi^2 \ell_{\text{eff}}^2} \label{kappaeff2}
\end{equation}
Contrary to the one-to-one correspondence observed in TTG, the effective conductivity inferred by TDTR is thus determined by a broad sampling of the generalised conductivity $\kappa^{\ast}(\xi)$. Kernel components with high spatial frequency $\xi \ell \gg 1$ are strongly suppressed and contribute little to the overall result, in accordance with the fact that the majority of thermal energy is contained within a few thermal penetration lengths. Expressing that $\ell_{\text{eff}} = \sqrt{\kappa_{\text{eff}}/(\pi C_0 f_{\text{mod}})}$ within the modified Fourier framework finally leads to an integral equation for the desired $\kappa_{\text{eff}}$:
\begin{equation}
\int \limits_{0}^{\infty} \frac{\kappa^{\ast}(\xi) \mathrm{d}\xi}{\pi C_0 f_{\text{mod}} + \kappa_{\text{eff}} \,\xi^2} - \sqrt{\frac{\pi \kappa_{\text{eff}}}{4 C_0 f_{\text{mod}}}} = 0 \label{integraleq}
\end{equation}
For pure L\'evy transport $\kappa^{\ast}(\xi) \sim 1/|\xi|^{2-\alpha}$ in semiconductor alloys, (\ref{integraleq}) can be solved fully analytically and produces $\kappa_{\text{eff}} \sim f_{\text{mod}}^{-(2/\alpha-1)}$ in accordance with experimental measurements \cite{alloys-part1,alloys-part2} on InGaAs and SiGe. More complicated $\kappa^{\ast}$ kernels can be easily treated numerically by solving (\ref{integraleq}) with a simple bisection algorithm with Gaussian quadrature for the $\xi$ integral.
\section{CONCLUSIONS}
In summary, we have presented a nonlocal framework for microscale heat conduction in semiconductor materials. Contrary to previous work, our formalism is applicable across the entire ballistic-diffusive spectrum, accommodates arbitrary phonon dispersions, and can be easily compared to actual measurement data through synergy with stocastic theory. We obtained a fully analytical solution for the spatial flux memory in tempered L\'evy transport. This is valuable to quasiballistic heat flow in semiconductor alloys as well as river sediment transport and solute chemical reactions. Our formalism offers interesting potential for improved compact thermal modeling and can serve as a generic framework to derive and study constitutive laws in a wide range of nondiffusive transport applications.
\section*{ACKNOWLEDGEMENTS}
A.S. acknowledges Purdue University start up funds that supported the research conducted by B.V.
%
\end{document}